\documentclass[11pt, authoryear]{llncs}

\usepackage[utf8]{inputenc} % allow utf-8 input
\usepackage[T1]{fontenc}    % use 8-bit T1 fonts
\usepackage{hyperref}       % hyperlinks
\usepackage{url}            % simple URL typesetting
\usepackage{booktabs}       % professional-quality tables
\usepackage{amsfonts}       % blackboard math symbols
\usepackage{nicefrac}       % compact symbols for 1/2, etc.
\usepackage{microtype}      % microtypography
\usepackage{graphicx}
\usepackage{siunitx}
\usepackage[titletoc,title]{appendix}
\usepackage{natbib}
\usepackage{amsfonts, amsmath}
\usepackage[algo2e, ruled, vlined]{algorithm2e}
\usepackage{xspace}
\usepackage{varwidth}
\usepackage{tikz}
\usepackage{bm}
\usepackage{fullpage}
\usepackage{enumerate}

\title{Are All Experts Equally Good? A Study of Analyst Earnings Estimates}

\author{Amir Ban\inst{1} \and Yishay Mansour\inst{2}\
\institute{Faculty of Industrial Engineering and Management, Technion, Israel Institute of Technology \email{amirban@netvision.net.il} \and Blavatnik School of Computer Science, Tel Aviv University \email{mansour.yishay@gmail.com}}}

\begin{document}

\maketitle

\begin{abstract}
We investigate whether experts possess differential
expertise when making predictions. We note that this would make it possible
to aggregate multiple predictions into a result that is
more accurate than their consensus average, and that the
improvement prospects grow with the amount of differentiation. Turning this argument on its head,
we show how differentiation can be measured by how much weighted aggregation improves on simple
averaging.
Taking stock-market analysts as experts in their domain, we do a
retrospective study using historical quarterly earnings 
forecasts and actual results for large publicly traded companies. 
We use it to shed new light on the \cite{sinha1997re} result, showing that analysts indeed possess individual expertise, but that their differentiation is modest. 
On the other hand, they have significant individual bias. 
Together, these enable a 20\%-30\% accuracy improvement over consensus average.
\end{abstract}

\section{Introduction}

Do experts have differential expertise? That is, can we grade and
rank experts by the quality of their expertise? To put these
questions in context, an expert is someone who knows about a domain
more than is commonly known. Often, this manifests itself in being able to predict
the outcome of a future event more accurately than laymen. The expert's level of
expertise is his expected accuracy, which, assuming the expertise is
real, may be learned from his past record of predictions. If there
is no correlation between an expert's past record and how well he
will do in his next prediction, the expert's results are random, and
there is no expertise.

The alternative to the existence of differential expertise is that
all experts (on a particular domain) have the same expected
accuracy. This can plausibly arise when all experts share a successful model (e.g., in football, that there is a home team advantage, 
and that the last 3-months win-lose record is a good indicator of future success), 
while any differences among them stem from random noise (for example, holding unfounded beliefs, such as that a particular team performs better in
large stadiums).

If differential expertise does not exist, the best way to
advantageously aggregate predictions of several experts, is to
average them. This reduces the noise element (variance) within
individual predictions, highlighting the common signal. On
the other hand, if experts do have differential expertise, then
averaging predictions, while usually still advantageous, is sub-optimal: A
weighted average, in which the better experts are given more
weight than their colleagues, should, in expectation, produce a
more accurate aggregate prediction.

We turn this argument on its head in order to measure the amount of differentiation in expertise:
Informally, the bigger the differentiation, the more unequal will be the weights assigned to experts, and
the larger the accuracy improvement (compared to unweighted aggregate) prospects of the prediction.
We {\em define} the amount of differentiation as the accuracy improvement that is achievable by weighting. To make this definition practical
and precise, (the best weighting method may be unknown), we restrict ourselves to weights calculated by a simple and reasonable algorithm,
whose results, if not optimal, are expected to be at least in the same order of magnitude. This will be sufficient for us to make {\em qualitative}
judgements about expert differentiation (e.g. ``Differentiation is large / modest / insignificant'').

We conducted an empirical study of this question, using earnings of
publicly-traded companies as the domain of predictions. The experts
here are stock-market analysts, and the outcomes are announced by
the companies on a quarterly or annual basis, on dates set well in
advance. There are several advantages to choosing this domain:
First, databases with historical predictions and results are
available for the past several decades. Second, analyst predictions,
and especially the ``surprise'', the difference between analyst
consensus\footnote{The consensus is defined on the NASDAQ website as
``The average EPS estimates for the company from the universe of
analysts reporting for the company \ldots''} and actual results, are
known to be major drivers of stock prices, so that these predictions
attract much interest from investors, and, due to the large amount
of money at stake, can be assumed to have best effort invested in
them. Third, much academic work has been done in the past half
century on the accuracy of analysts and what affects it.

To answer our question about expert differentiation we employ
algorithms that combine a weighted consensus estimate from individual predictions. The algorithms are not presented
as innovative or optimal, (indeed, the major part of their improvement is achieved with an algorithmically-trivial bias correction),
but are competent enough to show the order-of-magnitude improvement achievable.

The contribution of our work is in three areas:
\begin{enumerate}
\item We clarify the \cite{sinha1997re} result, described below, that found that analysts differ in accuracy, but did not quantify the size of this differentiation. We confirm the existence of accuracy differentiation, but show that it is rather modest, and of doubtful economic significance. In the process, we make the first aggregation of analyst estimates. The absence of published aggregation attempts in the curious and motivated academic community following Wall Street is puzzling. Lack of significant results may be the explanation, or perhaps secretiveness. 
\item The technique of devising improved weighted estimates is suggested as
a definite test for the existence and magnitude of differentiation among experts, and
the sensitivity of its results to changes of parameters is suggested as a method for model selection. The superiority of this method over statistical tests of significance is demonstrated. This is applicable to other fields where
multiple predictions are made: Sports competitions, election results, weather and climate forecasting,
revenue forecasting for new movies / books / albums, and more.
\item Despite the minor differentiation in expertise among analysts, we nevertheless do achieve a significant average improvement of 20\%-30\% in forecast accuracy over consensus. The major part of this improvement comes from an unexpected quarter: Systematic and individual analyst bias. While a general pessimistic bias is well-known (see below), we find a significantly larger {\em individual} analyst bias. This discovery was an unanticipated part of our work. The literature has struggled to explain the general bias of analysts (since self-calibration is the easiest part of prediction). The discovery of significant individual bias may shed light on this.
\end{enumerate}

\subsection{Related Literature}
\label{related}

In \cite{goel2010prediction}, the authors investigate the advantage
of market mechanisms over ``conventional methods of prediction,
namely polls and statistical models'' in several domains, and find a
``surprisingly small'' advantage to the markets, and ``remarkably
steep diminishing returns to information, with nearly all the
predictive power captured by only two or three parameters''.

Though the study does not ask the same questions we ask, it supports
the conclusion that differential expertise is either non-existent or
weak, in the domains investigated. This is because, as the authors
note, ``theoretical and empirical evidence suggests that markets do
often outperform alternative mechanisms''. If markets do not
strongly outperform ``two or three parameters'' models, then perhaps
nothing does.

A vast literature has analyzed analyst performance and the factors
affecting it for the past half century. For a survey, see, e.g.,
\cite{ramnath2008financial}. The consensus regarding differential
accuracy of analysts has gone through changes. \cite{o1990forecast}
studied the statistical relationship of analyst accuracy to the
identity of the analyst, the firm, and the year, and in regards to
the analyst concluded ``Overall, I find no systematic differences in
forecast accuracy across individuals''.

However, \cite{sinha1997re} repeated O'Brien's work and
overturned its conclusion by adding another element of variation.
Paraphrasing, they note that analyst forecasts are heavily
influenced by events, or shocks, which also causes them to change in
near-unison. When this effect, modeled as a time series, is added to
O'Brien's independent variables, the
identity of the analyst re-emerges as significantly affecting
forecast accuracy.

{\em Despite much further work on analyst accuracy and the factors affecting it, and despite suggestions to clarify its results (e.g., in  \cite{ramnath2008financial}), \cite{sinha1997re} remains the latest word on whether forecast accuracy differs among analysts. No later work attempts to quantify the differentiation}.

O'Brien, Sinha, and almost the entire literature on analyst earnings predictions aim to prove or disprove an underlying relationship to various attributes using statistical tests of significance. We consider tests of significance as inadequate for proving differential expertise, first, because statistical significance does not imply economic significance, and second, because it attaches no numerical magnitude to the differentiation. As opposed to the cited works, we develop a methodology of {\em weighting} to measure differentiation of expertise, then use its results to draw conclusions.

As noted, differences in expertise should make it possible to cast more accurate predictions by suitable weighting. Indeed, we argue that the ability to cast a significantly better prediction is a better test of significance, since, while clearly if a relationship is statistically insignificant it cannot lead to better predictions by weighting, the converse is not true: A statistically-significant relationship may lead to only negligibly better predictions through weighting.

Notwithstanding, with the exception of two works to be presently cited, we have not found any attempt to leverage proved relationships into more accurate consensus predictions. The survey \cite{ramnath2008financial} cites no such work, and merely suggests it for future research.

The latter of the two works, \cite{slavin2007aggregating}, (Slavin, henceforth), is based on, and improves on the earlier one \cite{brown2001profiting}, (BM, henceforth). Slavin claims to reduce forecast errors by 29\%, better than the 16\% achieved by BM. BM predict individual analyst accuracy using characteristics such as forecast recency, forecast frequency, brokerage reputation, experience, and the number of companies and industries covered. All these had been shown by even-earlier studies to be correlated with analyst accuracy. They then combine forecasts using weights that are based on each analyst's projected accuracy ranked relative to their peers. Slavin improves on their methodology with two steps. First, he uses past accuracy (in addition to their six factors) to predict individual analyst accuracy. Second, his forecast weight accounts for the magnitude of analyst accuracy, whereas BM use only ordinal ranking.

We use a modified version of their methodology in our present work, to model the individual expertise of analysts. (We supplement it with a model of individual bias). For us, it achieves only a fraction of the improvement reported above. There are many important differences between our data and implementation, and theirs, so this cannot be considered a reexamination of their studies. Nevertheless, we believe that the difference is too large to be explained by differences in details. If Slavin and BM ever produced results as good as they reported, then they no longer do, with more recent data. The reason we find their framework nevertheless useful is:

\begin{itemize}
\item The methodology, as described, is free of errors, such as inadvertently ``peeking'' into the future.
\item It uses most characteristics reported in the literature to be significantly correlated with analyst accuracy.
\item It seems unlikely that a different algorithm can gain much more from the same data, so what it achieves is a first-order approximation to what is achievable.
\end{itemize}

Better aggregation algorithms are probably available, for example \cite{raykar2009supervised}, \cite{raykar2010learning}. However, (i) they do not use domain-specific data. As noted, the literature on analyst performance has proven several attributes as significantly correlated with analyst accuracy. We cannot omit such proven attributes from our aggregation methods without putting the relevance of our work into question. {\em Inter alia}, our methodology will show that some of these attributes in fact have insignificant marginal impact. And (ii) As we argue later, an aggregation algorithm would need to be implausibly an order-of-magnitude better to change our qualitative conclusions.

In regards to our findings on analyst bias: Surprisingly, over the last three to four decades there has been a
consistent drift from slightly more negative surprises (i.e., analyst optimism) to
overwhelmingly positive surprises (i.e., analyst pessimism, see \cite{Brown2001} and
\cite{Matsumoto2002} for a documentation of the drift). The academic
financial research has struggled to explain the sources  for
this drift. Possible explanations include analysts' conflict of
interest (\cite{ChanKaLa2007}), management avoidance of negative
surprises (\cite{Matsumoto2002}), difference among various countries
(\cite{BrownHi2001}), and more.
Our findings are in a similar spirit, where the positive surprises
outnumber the negative ones. 

%\cite{dreman1995analyst,das1998earnings} for earlier works showing
%analysts optimism resulting in negativesurprises

%[[YM:  There are many additional works (in the dropbox) and it seems
%that over the decades the balance drifted from negative surprise to
%positive surprise. ]]
%
%Most studies of earning predictions find that predictions tend to be
%optimistic. \cite{dreman1995analyst} find that estimates
%overestimate actual earnings by an average of 3.6\%.
%\cite{das1998earnings}, among others, believe that the reason for
%analyst optimism is the need to curry favor with firm managers, and
%say ``Absent such incentive considerations, we would not expect
%earnings forecasts to be systematically biased. Forecasts could vary
%in accuracy across firms, but inaccuracy per se does not translate
%into a bias''.

%Our findings here, too, are sharply different.

\subsection{Summary of Results}

We describe an algorithm that calculates an improved consensus prediction out of the set of all analyst predictions preceding a quarterly earnings announcement, and we run it for all quarters in 2010-2016 for all firms who compose two well-known indices: The S\&P-500 index, and the NASDAQ-100 index.

We use the resulting accuracy of this improved consensus as a benchmark for significance, eschewing statistical tests of significance. This is because we note that, often, attributes that test positively for significance nevertheless have insignificant impact on our consensus prediction.

Measuring accuracy by the criterion used by Slavin and BM, the median improvement in surprise, we find an improvement of 21\% in the S\&P-500 group and 28\% in the NASDAQ-100 group.

Our conclusion, though, is that analyst differential expertise is modest. This is because the major part of the improvement stems from modeling individual bias. The rest, and much smaller part, comes from modeling individual expertise and other attributes. Modeling individual bias {\em only} improves estimates by 15\% for the S\&P-500 group and 22\% for the NASDAQ-100 group. Modeling individual expertise {\em only} (the Slavin / BM methodology) improves the S\&P-500 group by a mere 3\%, and the NASDAQ-100 group, by 5\%.

We find that bias is best modeled as a function of both firm and analyst. This yields much better results than correcting for a general bias, and is also superior to modeling bias as dependent on firm only, or on analyst only, or on any weighted combination of both. The conclusion is that analysts have a consistent {\em individual}  bias, quite apart from the general pessimistic bias displayed by the aggregate of analyst opinions.

If analyst identities are replaced by their institutions (brokerages), accuracy decreases. This indicates that individual bias and expertise are related to the analyst rather than to the employing institution.

%We found that predictions are markedly pessimistic, not optimistic. Only 30\% of S\&P-500 quarterly reports, and 28\% of NASDAQ-100 quarterly reports, have negative surprise.

\subsection{Organization of this Paper}

The rest of this paper is organized as follows. In Section \ref{data} \textsc{Data} we describe the data used. In Section \ref{methods} \textsc{Methods} we describe the methods and algorithms used. Results are presented in Section \ref{results} \textsc{Results}. In Section \ref{discussion} \textsc{Discussion} we summarize and offer concluding remarks.

\section{Data}
\label{data}

The data for this study were obtained from the Thomson Financial's Institutional Broker Estimate System (I/B/E/S), available through Wharton Research Data Services. Specifically the databases used were Detail History / Detail, Detail History / Actuals and Summary History / Surprise History.

Data was collected for two stock symbol lists\footnote{Companies occasionally enter or leave these symbol lists. Note that while we use the lists as of 2016, we use the record of the component companies over all the study history, whether or not they were part of the index at the time.}:
\begin{enumerate}
\item The S\&P-500 index, as of 2016 (504 symbols).
\item The NASDAQ-100 index, as of 2016 (103 symbols).
\end{enumerate}

Data was gathered on analyst estimates and actual results for all earnings announcements made between 2004 and 2016. We calculated our improved consensus prediction only for results announced 2010-2016, since we wanted it to rely on at least 6 years of analyst performance history.

For each quarterly earnings announcement, the actual earnings were taken from both the Actuals and the Surprise History datasets. Earnings where the two values did not match were discarded. To exclude probable errors, earnings with absolute surprise of more than 50 cents per share were also discarded. This caused less than 5\% of the reports to be excluded. Also, only earnings announcements which at least 8 different analysts estimated were considered. Analyst estimates made within a year of the earnings announcement (specifically, having I/B/E/S field FPI = 6, 7, 8 or 9), and no later than 48 hours\footnote{Later estimates are suspected to be errors in data.} before the announcement, were considered. If an analyst issued several estimates for the same earnings announcement, only the last one was considered. Also excluded were predictions where the analyst had no record of previously estimating the firm. This guarantees that data exists to estimate past accuracy.

\section{Methods}
\label{methods}

For readability, we describe our methods verbally. For added clarity, the reader will find formulas in the Appendix.

We number the periods (quarters) under study $1,2,...$ in chronological order. Let $ACTUAL_{j,t}$ be the actual earnings of firm $j$ at period $t$, and, if analyst $i$ made a prediction of it, let $PREDICT_{i,j,t}$ be that prediction.

The signed prediction error of analyst $i$ of firm $j$ for period $t$, $ERR_{i,j,t}$, is the difference between the analyst prediction for the period and the actual earnings.

The experience at $t$ of analyst $i$ with firm $j$ is the number of periods prior to $t$ that the analyst made a prediction for the firm.

\subsection{Bias Calculation}

Bias is tracked for each analyst-firm combination. $BIAS_{i,j,t}$ is the bias of analyst $i$ for firm $j$ at time $t$. It is defined as the average of the prediction error $ERR_{i,j,\tau}$ for all periods previous to $t$ where the analyst made a prediction for firm $j$. If there are no such periods, the bias is defined to be $0$.

The bias-adjusted prediction of an analyst is his prediction, minus his bias at the current period. Similarly $AE_{i,j,t}$, the adjusted prediction error is the prediction error minus the current bias.

\subsection{Modeling Prediction Errors}
\label{model}

Following \cite{slavin2007aggregating} (see Section \ref{related} \textsc{Related Literature}, where we explained our rationale for following that framework), we best-fit a multiple linear model to prediction data in every calendar quarter in our study (2010-2016), then use the resulting model to predict individual prediction errors in the {\em next} quarter. These predicted errors are then used to attach weights to individual analyst predictions, (the smaller the predicted error, the larger the weight). Our improved consensus estimate is the resulting weighted average of analyst estimates.

The reader should bear in mind our intended conclusion: The results of this aggregation algorithm (without bias adjustment) are {\em not significant}, achieving, as shown below, a 3\%-5\% accuracy improvement over simple consensus. This is less than 1/5 of the total improvement achieved, the rest attributable to bias adjustment. The algorithm is not presented as optimal, but as ``good enough'', and, according to existing literature, using all relevant data. Better algorithms might yield better results,  but not enough to change the conclusion, which is that expertise differentiation among analysts is modest.

There are some differences between our model and Slavin's, the major one being that we adjust all predictions for individual bias. Other differences are listed in the Appendix.

\subsubsection{Variables}
\label{variables}

The dependent variable, $AAE_{i,j,t}$, is the absolute bias-adjusted prediction error of analyst $i$ of firm $j$ for period $t$.

The independent variables are:
\begin{enumerate}
\item $AGE_{i,j,t}$: The number of days from the prediction to the earnings announcement.
\item $FREQ_{i,j,t}$: The number of predictions this analyst has made for the current quarter.
\item $NCOS_{i,j,t}$: The number of different firms this analyst is covering in this quarter.
\item $TOP10_{i,j,t}$: $1$ if the institution for which this analyst works is in the top decile of institutions, $0$ if not, where institutions are ranked by the number of analysts reporting for them in the current quarter.
\item $EXP_{i,j,t}$: The number of times this analyst previously reported on this firm.
\item $MAE_{i,j,t}$: Mean of $AAE_{i,j,\tau}$ in all this analyst's previous predictions for this firm. (Earnings for which there are no such previous predictions are excluded).
\end{enumerate}

The variables $DAAE_{i,j,t}$, $DAGE_{i,j,t}$, $DFREQ_{i,j,t}$, $DNCOS_{i,j,t}$, $DTOP10_{i,j,t}$, $DEXP_{i,j,t}$ and $DMAE_{i,j,t}$ are versions of the above, normalized and scaled to their all-analyst average: e.g. $DAGE_{i,j,t} = AGE_{i,j,t} / AGE_{j,t} - 1$ ($AGE_{j,t}$ stands for the all-analyst average of $AGE_{i,j,t}$).

\subsubsection{Linear Model}
\label{linear-model}

Our multiple linear model is
\begin{eqnarray}
\label{best-fit}
DAAE_{i,j,t}^* = & \beta_{1,t }DAGE_{i,j,t} +  \beta_{2,t} DFREQ_{i,j,t}  \nonumber \\
&+ \beta_{3,t} DNCOS_{i,j,t} + \beta_{4,t} DTOP10_{i,j,t}  \nonumber \\
&+ \beta_{5,t} DEXP_{i,j,t}  + \beta_{6,t} DMAE_{i,j,t} + \epsilon_{i,j,t}
\end{eqnarray}

For each period $t$, multiple linear regression is used to find $\beta_{k,t}, k= 1, \ldots, 6$ that best-fit this model, minimizing the sum of squares of the errors $\epsilon_{i,j,t}$: $$\sum_{i,j} (DAAE_{i,j,t}^* - DAAE_{i,j,t})^2$$

A prediction of $DAAE_{i,j,t}$, marked $\widehat{DAAE}_{i,j,t}$, can then be calculated from {\em current} attributes and the {\em previous} period's model, as follows
\begin{eqnarray}
\label{predict-error}
\widehat{DAAE}_{i,j,t} = & \beta_{1,t -1}DAGE_{i,j,t} +  \beta_{2,t-1} DFREQ_{i,j,t}  \nonumber \\
& +  \beta_{3,t-1} DNCOS_{i,j,t} +  \beta_{4,t-1} DTOP10_{i,j,t} \nonumber \\
&+  \beta_{5,t-1} DEXP_{i,j,t}  + \beta_{6,t-1} DMAE_{i,j,t}
\end{eqnarray}

\subsubsection{Calculating a Weighted Consensus}
\label{weights}

Our consensus prediction $PREDICTION_{j,t}$ is a weighted average of analyst $i$ predictions for firm $j$ in period $t$. The weights are set as follows:

For each analyst $i$ reporting on firm $j$ in period $t > 1$, the
predicted error is calculated by (\ref{predict-error}), using model \eqref{best-fit} that was best-fit for the {\em previous} period. Then the
average (over analysts) predicted error $\widehat{DAAE}_{j,t}$ is
calculated. Analysts whose predicted error exceeds the all-analyst average
are given zero weight, while others are given
weight $ (\widehat{DAAE}_{j,t} - \widehat{DAAE}_{i,j,t})^r$. The result is not very sensitive to $r$. We use $r=1.2$, which we find fit the data better than Slavin's $r=2$ (see Table \ref{result-table}).

\section{Results}
\label{results}

\subsection{Statistics Used}

Our results compare how our weighted consensus prediction $PREDICTION_{j,t}$ measures against the simple-average consensus $CONSENSUS_{j,t}$. The simple-average consensus is widely used and reported by the financial community and popular websites, and so is the natural yardstick.

We use three different statistics to evaluate the total improvement of the result for a given list of firms.
\begin{itemize}
\item \textsc{median}: Defined as the median, taken over all firms $j$ in the list and all periods $t$ in the target range, of $SURPIMPR_{j,t}$, the fractional improvement in the absolute surprise.

\item \textsc{average}: This is the ratio of the sum of all improved absolute surprises divided by the sum of the original absolute surprises.

\item \textsc{trend}: This is $1$ minus the slope of the linear regression of the improved surprise $PREDICTION_{j,t} - ACTUAL_{j,t}$ vs. the original surprise $CONSENSUS_{j,t} - ACTUAL_{j,t}$ (see Figure \ref{trend}).

\end{itemize}

The \textsc{median} statistic is not sensitive to the magnitude of the stock price. On
the other hand the \textsc{average} statistic might over-emphasize high stock prices,
assuming earnings and surprises are in proportion. The \textsc{median} statistic over-emphasizes announcements with low surprise (for zero surprise $SURPIMPR_{j,t}$ is infinite), but,
being ordinal, the effect is not excessive. On the other hand the \textsc{average} statistic, in effect, weighs
announcements by surprise, so under-emphasizes announcements with low surprise.

The \textsc{trend} statistic is the only one that takes into account the sign of the surprise. See Figure \ref{trend}. However, we consider this statistic to be less trustworthy than the other two, and do not use it for mode sensitivity analysis. This is because it is sensitive to outliers, and is meaningless when the correlation coefficient is low.

\subsection{The Results}

\begin{table*}[tb]
  \caption{Statistics on Earnings Reports}
  \label{stats-table}
  \centering
 \begin{tabular}{lrr}
\toprule
& S\&P-500 & NASDAQ-100 \\
\midrule
Number of symbols & 504 & 103 \\
Number of reports & 10,776 & 2,126 \\
Number of predictions & 434,255 & 102,217 \\
Number of analysts & 5,711 & 2,701 \\
Average absolute surprise (cents per share) & 7 & 7.2 \\
Median absolute surprise (cents per share) & 3.8 & 3.2 \\
Reports with negative surprise & 30.3\% & 27.8\% \\
Reports with actual in prediction range & 75.4\% & 73.9\% \\
\bottomrule
\end{tabular}
\end{table*}

\begin{table*}[tb]
  \caption{Improvement Results for Method and Variations}
  \label{result-table}
  \centering
 \begin{tabular}{lrrrrrr}
\toprule
&
\multicolumn{3}{c@{\quad}}{S\&P-500}
&
\multicolumn{3}{c}{NASDAQ-100} \\\cmidrule(r){2-4}\cmidrule(l){5-7}
Mode &\textsc{median} & \textsc{average} & \textsc{trend}  &\textsc{median} & \textsc{average} & \textsc{trend}   \\
\midrule
Full mode & 20.9\% & 16.7\% & 24.6\% & 27.7\% & 23.9\% & 28.9\% \\
\\
Without individual expertise & 14.7\% & 9.9\% && 22.2\% & 12.8\% \\
Without individual bias & 2.6\% & 4.8\% && 5.2\% & 12.1\%\\
\\
Without $AGE$ & 19.0\% & 14.5\% && 26.0\% & 20.2\% \\
Without $FREQ$ & 20.8\% & 16.6\% && 27.8\% & 24.0\% \\
Without $TOP10$ & 20.9\% & 16.7\% && 27.9\% & 23.9\% \\
Without $NCOM$ & 20.9\% & 16.7\% && 27.1\% & 24.0\% \\
Without $EXP$ & 20.5\% & 16.6\% && 27.4\% & 24.0\%\\
Without $MAE$ & 17.9\% & 14.3\% && 26.2\% & 22.6\% \\
Without scaling of variables & 19.4\% & 15.5\% && 27.2\% & 22.3\%  \\
\\
General bias  &14.5\% & 11.4\% && 19.8\% & 17.5\% \\
Bias based on firm only &16.9\% & 13.1\% && 25.8\% & 22.1\% \\
Bias based on analyst only & 18.1\% & 14.7\% && 24.8\% & 20.8\% \\
Bias weighted half-firm, half-analyst & 19.5\% & 15.7\% && 25.8\% & 23.0\% \\
\\
Use institution instead of analyst id & 19.8\% & 16.2\% && 27.6\% & 21.3\% \\
Use Slavin's exponent in Section \ref{weights} & 20.2\% & 16.0\% && 27.0\% & 24.2\% \\
Estimates up to 30 days before earnings & 16.1\% & 12.1\% && 23.8\% & 19.1\% \\
Estimates up to 60 days before earnings & 13.0\% & 9.1\% && 21.4\% & 17.6\% \\
\\
Closest analyst & 90.0\% & 80.0\% && 93.2\% & 85.5\% \\
Closest analyst without bias correction & 96.4\% & 70.8\% && 100.0\% & 76.6\% \\
\bottomrule
\end{tabular}
\end{table*}

\begin{figure*}[!tbp]
  \centering
  \begin{minipage}[b]{0.49\textwidth}
    \includegraphics[height=0.3\textheight]{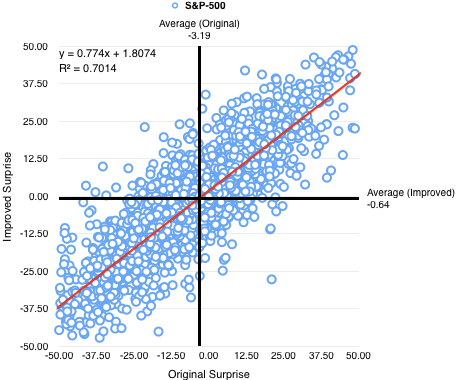}
  \end{minipage}
  \hfill
  \begin{minipage}[b]{0.49\textwidth}
    \includegraphics[height=0.3\textheight]{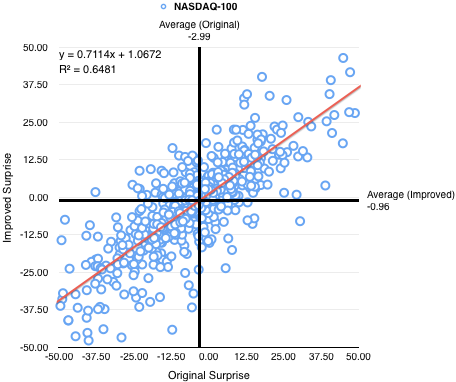}
  \end{minipage}
\caption[Original and improved prediction errors with trend-line]{Original and improved prediction errors with trend-line}
\label{trend}
\end{figure*}

Table \ref{stats-table} lists several statistics about the earnings reports on which we ran this study.

Results are tabulated in Table \ref{result-table}. The three statistics are shown for the two firm groups. The first row has the results for the full mode, as described above. The other rows are a sensitivity analysis, showing results for various variations and omissions, as described in the {\em Mode} column.

With hundreds of earnings reports per period (see Table \ref{stats-table}), the number of adjustable parameters used in our model (6) and for model selection (see Table \ref{result-table}) is small in comparison. No overfitting takes place.

Figure \ref{trend} shows all predictions used in two graphs, one per symbol list. Each prediction used is represented by a blue circle. Its horizontal coordinate is the original surprise, and its vertical coordinate is the surprise improved by the model. The axes are placed at the average of the respective surprise. Note that their values are negative, reflecting the tendency of analysts to underestimate earnings. The red trend line is the result of simple linear regression on the prediction points, with the linear relationship and R-squared value in the upper left-hand corner. The linear regression is used to compute the \textsc{trend} statistic. The fact that the trend-line slope is less than 1 (angle less than \ang{45}) shows the general improvement of predictions by the model. The \textsc{trend} statistic is read as 1 minus the slope coefficient of the regression.

In general, we observe that results for the NASDAQ-100 group are consistently better than for the S\&P-500 group, indicating that for this group, differentiation among analysts is more significant. We also note that for all modes that include individual bias, the \textsc{median} statistic is higher than the \textsc{average}, indicating the presence of low-improvement outliers. The \textsc{trend} statistic yields the highest results.

The rows {\em Without individual bias} and {\em Without individual expertise} demonstrate that the major part of the method's improvement derives from modeling individual bias. Modeling individual expertise, i.e. Section \ref{model} \textsc{Model}, on its own, yields single-digit percentages in the \textsc{median} statistic,  less than one-fifth of the total.

Modes starting {\em Without $AGE$} show the effect of omitting each of the independent variables in Section \ref{variables} \textsc{Variables}, demonstrating its marginal contribution to accuracy. The effect is insignificant for $FREQ$, $TOP10$, $NCOM$ and $EXP$, showing that, by our criteria, these variables do not significantly affect the prediction accuracy. $AGE$ and $MAE$ are significant, as is the normalizing and scaling of the variables (Section \ref{variables} \textsc{Variables}).

Several modes check variations on correcting for bias based on analyst-firm combination. They show that correcting based on general (not analyst or firm-specific) bias is decidedly inferior. They also show that correcting based on analyst-only, on firm-only, or even using an average of the per-analyst bias and the per-firm bias, are all inferior. This is despite the fact that analyst-firm combinations have fewer samples, so that the related bias, calculated by averaging those, is more noisy.

If instead of modeling bias and expertise for individual analysts, we do the same for individual brokerages (the employers of the analysts), the results are inferior.

Cutting off predictions 30 or 60 days before earnings announcements results in significantly less improvement. This indicates that the differentiation among analysts is more significant for forecasts nearer earnings.

Finally, we check how good a ``prediction'' we could get if we followed the best analyst, in hindsight. This is tantamount to asking how close the closest analyst came to the actual result. The answer we find is, very close. Indeed, the $100\%$ \textsc{median} result for the NASDAQ-100 group means that in more than 50\% of these earnings announcements, there was an analyst who got it exactly right. In view of the other results, we find this result mildly surprising. It seems to show that our difficulty in predicting earnings is not so much because all analysts are far from the truth, but because we fail to identify in advance those who got it right. A related result shown in Table \ref{stats-table} is that about $75\%$ of actuals fall between the lowest and highest analyst prediction.

\section{Discussion}
\label{discussion}

\subsection{Conclusions}

We performed an empirical study of aggregating analyst predictions to an improved-accuracy consensus, and used its results to show that, while analysts have differential expertise, the differentiation is not large. We described in detail a methodology that shows an economically significant improvement of 20\%-30\% in accuracy over standard consensus methods. A side benefit of this methodology is that it allows us to gauge in a novel fashion the significance, or insignificance, of many parameters that are hypothesized to affect analyst accuracy. We use it to show, {\em inter alia}, that several analyst parameters considered significant to accuracy, are in fact of low marginal significance.

We find that analysts, and even more notably, analyst-firm combinations, have both individual expertise and individual bias. The individual bias part is surprisingly large, and economically important in its own right. We find it is significantly bigger than the general (not specific to an analyst or firm) bias, and much more significant than the individual expertise, which may not be economically significant on its own.

When analyst identities are replaced by those of their institutions, the importance of the individual element is reduced, indicating that differentiation is mainly among individuals rather than among their institutions.

\subsection{Beyond Earnings}

Looking beyond our subject of earnings estimates, our study, though confirming the existence of differential expertise, strengthens and confirms \cite{goel2010prediction}'s observation (see Section \ref{related} \textsc{Related Literature}) of ``remarkably
steep diminishing returns to information''. If not for the substantial contribution of individual bias, we would be unable to leverage decades of research into analyst performance to more than a few percentage points of improved accuracy.

On the other hand, when including bias, we found significant total differentiation among experts. Would it not be correct to call individual bias (more precisely, a lack of it) a form of individual expertise? After all, an expert with little or no bias is more accurate than one with a large bias, all else being equal. The problem with doing that is that systematic bias is akin to lack of calibration in binary predictions (e.g. when a forecaster attaches a 70\% probability to all events that actually happen 60\% of the time). Being well-calibrated is often considered as separate from being {\em precise}, which, in binary predictions, means successfully attaching probabilities that are far from the non-committal 50\%.

There is a good practical reason for this distinction: Systematic bias (like lack of calibration) is easily observed and corrected for (even self-corrected for), unlike other sources of inaccuracy. Our discovery of widespread and significant bias in the well-followed universe of earnings predictions is therefore surprising.

\subsection{Other Domains}

The question we ask, ``Are all experts equally good?'', and our follow-up question, "And if not, how different are they?", is relevant to other domains of prediction. The implication, of what the potential is for improving predictions by aggregation, also applies.

This would be of great interest, for these domains in particular, and for the general question of what experts, in general, know, and what they do not. Our methodology as explained in this paper can provide answers given the data. We have looked, with the aim of continuing our investigation, but as yet have not found other domains with suitably available data. This would require:

\begin{itemize}
\item A domain of predictions, preferably of continuous outcomes.
\item An archive of prediction outcomes.
\item An archive of individual predictions of one of these outcomes, made by identifiable experts in the domain, whether they call themselves that way or not, whether they are individuals or institutions, and whether they see themselves as making predictions or not. (A bettor making a bet in a horse race, for example, would suffice).
\item Several individual predictions should be made of each outcome, preferably many.
\end{itemize}

Possibly there are sources that are open to insiders, and may be available to researchers through some agreement. We would be interested.

\bibliographystyle{ACM-Reference-Format-Journals}
\bibliography{earnings}


\begin{thebibliography}{00}

%%% the Web address.

\ifx \showCODEN    \undefined \def \showCODEN     #1{\unskip}     \fi
\ifx \showDOI      \undefined \def \showDOI       #1{{\tt DOI:}\penalty0{#1}\ }
  \fi
\ifx \showISBNx    \undefined \def \showISBNx     #1{\unskip}     \fi
\ifx \showISBNxiii \undefined \def \showISBNxiii  #1{\unskip}     \fi
\ifx \showISSN     \undefined \def \showISSN      #1{\unskip}     \fi
\ifx \showLCCN     \undefined \def \showLCCN      #1{\unskip}     \fi
\ifx \shownote     \undefined \def \shownote      #1{#1}          \fi
\ifx \showarticletitle \undefined \def \showarticletitle #1{#1}   \fi
\ifx \showURL      \undefined \def \showURL       #1{#1}          \fi

\bibitem[\protect\citeauthoryear{Brown and Mohammad}{Brown and
  Mohammad}{2001}]%
        {brown2001profiting}
{LD Brown} {and} {E Mohammad}. 2001.
\newblock {\em Profiting from predicting individual analyst earnings forecast
  accuracy}.
\newblock {T}echnical {R}eport. SSRN Working Paper.
\newblock


\bibitem[\protect\citeauthoryear{Brown}{Brown}{2001}]%
        {Brown2001}
{Lawrence~D. Brown}. 2001.
\newblock \showarticletitle{A Temporal Analysis of Earnings Surprises: Profits
  versus Losses}.
\newblock {\em Journal of Accounting Research\/} {39}, 2 (2001), 221--241.
\newblock


\bibitem[\protect\citeauthoryear{Brown and Ngo~Higgins}{Brown and
  Ngo~Higgins}{2001}]%
        {BrownHi2001}
{Lawrence~D. Brown} {and} {Huong Ngo~Higgins}. 2001.
\newblock \showarticletitle{Managing earnings surprises in the US versus 12
  other countries}.
\newblock {\em Journal of Accounting and Public Policy\/} {20}, 4-5 (2001),
  373--398.
\newblock
\showURL{%
\url{http://EconPapers.repec.org/RePEc:eee:jappol:v:20:y:2001:i:4-5:p:373-398}}


\bibitem[\protect\citeauthoryear{Chan, Karceski, and Lakonishok}{Chan
  et~al\mbox{.}}{2007}]%
        {ChanKaLa2007}
{Louis K.~C. Chan}, {Jason Karceski}, {and} {Josef Lakonishok}. 2007.
\newblock \showarticletitle{Analysts' Conflicts of Interest and Biases in
  Earnings Forecasts}.
\newblock {\em JOURNAL OF FINANCIAL AND QUANTITATIVE ANALYSIS\/} {42}, 4
  (2007), 893--914.
\newblock


\bibitem[\protect\citeauthoryear{Goel, Reeves, Watts, and Pennock}{Goel
  et~al\mbox{.}}{2010}]%
        {goel2010prediction}
{Sharad Goel}, {Daniel~M Reeves}, {Duncan~J Watts}, {and} {David~M Pennock}.
  2010.
\newblock \showarticletitle{Prediction without markets}. In {\em Proceedings of
  the 11th ACM conference on Electronic commerce}. ACM, 357--366.
\newblock


\bibitem[\protect\citeauthoryear{Matsumoto}{Matsumoto}{2002}]%
        {Matsumoto2002}
{D.~A. Matsumoto}. 2002.
\newblock \showarticletitle{Management's Incentives to Avoid Negative Earnings
  Surprises}.
\newblock {\em Accounting Review\/} {77}, 3 (2002), 483--514.
\newblock


\bibitem[\protect\citeauthoryear{O'Brien}{O'Brien}{1990}]%
        {o1990forecast}
{Patricia~C O'Brien}. 1990.
\newblock \showarticletitle{Forecast accuracy of individual analysts in nine
  industries}.
\newblock {\em Journal of Accounting Research\/} (1990), 286--304.
\newblock


\bibitem[\protect\citeauthoryear{Ramnath, Rock, and Shane}{Ramnath
  et~al\mbox{.}}{2008}]%
        {ramnath2008financial}
{Sundaresh Ramnath}, {Steve Rock}, {and} {Philip Shane}. 2008.
\newblock \showarticletitle{The financial analyst forecasting literature: A
  taxonomy with suggestions for further research}.
\newblock {\em International Journal of Forecasting\/} {24}, 1 (2008), 34--75.
\newblock


\bibitem[\protect\citeauthoryear{Raykar, Yu, Zhao, Jerebko, Florin, Valadez,
  Bogoni, and Moy}{Raykar et~al\mbox{.}}{2009}]%
        {raykar2009supervised}
{Vikas~C Raykar}, {Shipeng Yu}, {Linda~H Zhao}, {Anna Jerebko}, {Charles
  Florin}, {Gerardo~Hermosillo Valadez}, {Luca Bogoni}, {and} {Linda Moy}.
  2009.
\newblock \showarticletitle{Supervised learning from multiple experts: whom to
  trust when everyone lies a bit}. In {\em Proceedings of the 26th Annual
  international conference on machine learning}. ACM, 889--896.
\newblock


\bibitem[\protect\citeauthoryear{Raykar, Yu, Zhao, Valadez, Florin, Bogoni, and
  Moy}{Raykar et~al\mbox{.}}{2010}]%
        {raykar2010learning}
{Vikas~C Raykar}, {Shipeng Yu}, {Linda~H Zhao}, {Gerardo~Hermosillo Valadez},
  {Charles Florin}, {Luca Bogoni}, {and} {Linda Moy}. 2010.
\newblock \showarticletitle{Learning from crowds}.
\newblock {\em Journal of Machine Learning Research\/} {11}, Apr (2010),
  1297--1322.
\newblock


\bibitem[\protect\citeauthoryear{Sinha, Brown, and Das}{Sinha
  et~al\mbox{.}}{1997}]%
        {sinha1997re}
{Praveen Sinha}, {Lawrence~D Brown}, {and} {Somnath Das}. 1997.
\newblock \showarticletitle{A Re-Examination of Financial Analysts'
  Differential Earnings Forecast Accuracy}.
\newblock {\em Contemporary Accounting Research\/} {14}, 1 (1997), 1--42.
\newblock


\bibitem[\protect\citeauthoryear{Slavin}{Slavin}{2007}]%
        {slavin2007aggregating}
{Greg Slavin}. 2007.
\newblock \showarticletitle{Aggregating Earnings per Share Forecasts}.
\newblock  (2007).
\newblock


\end{thebibliography}

\newpage

\begin{appendix}

\section{Formulas}

Let $PERIODS_{i,j,t}$ be the set of periods up to $t$ for which analyst $i$ made a prediction for firm $j$.

\begin{equation}
ERR_{i,j,t} := PREDICT_{i,j,t} - ACTUAL_{j,t}
\end{equation}
\begin{equation}
\label{exp}
EXP_{i,j,t} := |PERIODS_{i,j,t}|
\end{equation}

If $EXP_{i,j,t} = 0$, we define $BIAS_{i,j,t} := 0$. Otherwise, it is
\begin{equation}
BIAS_{i,j,t} := \frac{\sum_{\tau \in PERIODS_{i,j,t}} ERR_{i,j,\tau}}{EXP_{i,j,t}}
\end{equation}

The prediction error adjusted for bias is
\begin{equation}
AE_{i,j,t} = ERR_{i,j,t} - BIAS_{i,j,t}
\end{equation}

\begin{equation}
\label{aae}
AAE_{i,j,t} := |AE_{i,j,t}|
\end{equation}

All dependent and independent variables are normalized and scaled to their all-analysts average:
\begin{eqnarray}
\label{scalefrom}
DAGE_{i,j,t} &:= (AGE_{i,j,t} - AGE_{j,t}) / AGE_{j,t} \\
DFREQ_{i,j,t} &:= (FREQ_{i,j,t} - FREQ_{j,t}) / FREQ_{j,t} \\
DNCOS_{i,j,t} &:= (NCOS_{i,j,t} - NCOS_{j,t}) / NCOS_{j,t} \\
DTOP10_{i,j,t} &:= (TOP10_{i,j,t} - TOP10_{j,t}) / TOP10_{j,t} \\
DEXP_{i,j,t} &:= (EXP_{i,j,t} - EXP_{j,t}) / EXP_{j,t} \\
\label{scaleto}
DMAE_{i,j,t} &:= (MAE_{i,j,t} - MAE_{j,t}) / MAE_{j,t} \\
\nonumber \\
DAAE_{i,j,t} &:= (AAE_{i,j,t} - AAE_{j,t}) / AAE_{j,t}
\end{eqnarray}

The weighted-average prediction is calculated using

\begin{eqnarray}
\label{weight}
w_{i,j,t} :=  \left\{
     \begin{array}{ll}
          0 & \widehat{DAAE}_{i,j,t} > \widehat{DAAE}_{j,t} \\
          (\widehat{DAAE}_{j,t} - \widehat{DAAE}_{i,j,t})^r & $otherwise$
     \end{array}
  \right.
\end{eqnarray}

\begin{equation}
\label{prediction}
PREDICTION_{j,t} := \frac{\sum_{i: t \in PERIOD_{i,j,t}} w_{i,j,t} PREDICT_{i,j,t}}{\sum_{i: t \in PERIODS_{i,j,t}} w_{i,j,t}}
\end{equation}

The standard, unweighted consensus is
\begin{equation}
\label{consensus}
CONSENSUS_{j,t} := \frac{\sum_{i: t \in PERIOD_{i,j,t}} PREDICT_{i,j,t}}{|\{i: t \in PERIODS_{i,j,t}\}|}
\end{equation}

The \textsc{median} statistic is the median of $SURPIMPR_{j,t}$
\begin{equation}
SURPIMPR_{j,t} := 1 - \frac{|PREDICTION_{j,t} - ACTUAL_{j,t}|}{|CONSENSUS_{j,t} - ACTUAL_{j,t}|}
\end{equation}

The \textsc{average} statistic is
\begin{equation}
AVGSTAT := \frac{\sum_j\sum_t|PREDICTION_{j,t} - ACTUAL_{j,t}|}{\sum_j\sum_t|CONSENSUS_{j,t} - ACTUAL_{j,t}|}
\end{equation}

\section{Differences from Slavin's Methods}
\label{differences}

For the record, the following are the major differences between Slavin's (and BM's) methods and ours (except for adjusting for bias, which is absent in Slavin). Slavin reported a much larger accuracy improvement than we do, but we are skeptical that the differences suffice to explain the divergence between our respective results. In addition, some of these changes make our method, more, not less accurate.

\begin{enumerate}
\item We consider quarterly rather than annual earnings announcements, and cover a different range of years. We consider only large, highly-traded firms.
\item Slavin discards all estimates made less than 30 days before the reported period end (which is about 2 months before announcement). We use all estimates made no later than 2 days before announcement.
\item One of Slavin's independent variables is the analyst's prediction error in the previous period, and therefore he discards all analyst predictions where no such previous prediction was made. Instead, we use the average of all prediction errors for the analyst, and discard only first predictions (per analyst-firm).
\item We do not include the $NIND$ (number of industries) independent variable, which is anyway described as having insignificant impact by Slavin.
\item We do not lag some independent variables by one period, as we find no reason to do so.
\item We use exponent $r = 1.2$ rather than $r = 2$ in \textit{Calculating a Weighted Consensus}.
\end{enumerate}

\end{appendix}
\end{document}